\documentclass[aps,preprint,superscriptaddress,longbibliography]{revtex4-1}
\usepackage{amsmath}
\usepackage{graphicx}
\usepackage{nicefrac}
\usepackage{subfigure}
\usepackage{subfig}

\begin{document}

\title{Effects of Nonlinear Decoherence on Halo Formation}

\author{S. D. Webb}
\email[]{swebb@txcorp.com}
\homepage[]{www.txcorp.com}
\affiliation{Tech-X Corporation, 5621 Arapahoe Ave. Suite A, Boulder, Colorado 80303, USA}

\author{D. L. Bruhwiler}
\altaffiliation[present address: ]{Center for Integrated Plasma Studies, University of Colorado, Boulder CO 80309}
\affiliation{Tech-X Corporation, 5621 Arapahoe Ave. Suite A, Boulder, Colorado 80303, USA}

\author{S. Nagaitsev}
\affiliation{Fermi National Accelerator Laboratory, Batavia, Illinois 60510, USA}

\author{V. Danilov}
\affiliation{Oak Ridge National Laboratory, Oak Ridge, Tennessee 37830, USA}

\author{A. Valishev}
\affiliation{Fermi National Accelerator Laboratory, Batavia, Illinois 60510, USA}

\author{D. T. Abell}
\affiliation{Tech-X Corporation, 5621 Arapahoe Ave. Suite A, Boulder, Colorado 80303, USA}

\author{A. Shishlo}
\affiliation{Oak Ridge National Laboratory, Oak Ridge, Tennessee 37830, USA}

\author{K. Danilov}
\affiliation{Tech-X Corporation, 5621 Arapahoe Ave. Suite A, Boulder, Colorado 80303, USA}
\affiliation{Physics Department, University of Colorado, Boulder, Colorado 80309-0390}

\author{J. R. Cary}
\affiliation{Tech-X Corporation, 5621 Arapahoe Ave. Suite A, Boulder, Colorado 80303, USA}
\affiliation{Physics Department, University of Colorado, Boulder, Colorado 80309-0390}

\date{\today}

\begin{abstract}
High intensity proton linacs and storage rings are central for the development of advanced neutron sources, extending the intensity frontier in high energy physics, as drivers for the production of pions in neutrino factories or muon colliders, and for the transmutation of radioactive waste. Such high intensity beams are not attainable using conventional linear lattices. It has been shown in the single particle limit that integrable nonlinear lattices permit much larger tune spreads than conventional linear lattices, which would mitigate many of the space charge restrictions that limit intensity. In this paper, we present numerical studies of space charge effects on a trial nonlinear lattice with intense bunches. We observe that these nonlinear lattices and their accompanying tune spreads strongly mitigate halo formation using a result from the particle-core model known to cause halo formation in linear lattices.
\end{abstract}


\maketitle

\section{Introduction}

High intensity beams have broad applications for high energy physics, neutron sources, and in the nuclear industry. For example, at Fermilab the Project X machine will deliver MW proton beams in the range of 3 to 120 GeV for the purposes of generating neutrinos for the proposed Long Baseline Neutrino Experiment, muons and kaons for the measurement of rare decays. It will also be used for the creation of exotic nuclei, and as a testbed for the transmutation of nuclear waste. For the intense beams desired for these purposes it is necessary to keep beam loss to a minimum to prevent activating the surrounding equipment. For example, in the 1.4 MW CW beam at SNS, it is necessary to keep beam loss to the pipe below 1 W/m. It is therefore crucial for these future applications that methods of mitigating intensity-dependent effects be developed.

A major limitation to the intensity of beams comes from the tune shifts resulting from transverse space charge~\cite{sac_68}. For a sufficient space charge driven tune shift, the bunch dynamics can fall on top of a resonance line and go unstable. However, linear lattices are desirable because their trajectories are integrable and therefore well understood. If a lattice could be designed which accepted large tune spreads while still having trajectories that are integrable or nearly integrable, the intensity limitations could be relaxed.

Kapchinskij and Vladimirskij provided the first study of the intensity limits on a linear lattice~\cite{kv_59}, developing equations for the envelope function of a beam which uniformly populates ellipses in any 2D projection of the phase space. Although unphysical, this so-called KV distribution is useful for theoretical and numerical studies. The resulting equations show that the beam envelope grows nonlinearly with beam current. Sacherer~\cite{sac_68} presented a thorough theoretical study of beam envelope functions assuming coasting beams with uniform transverse focusing, which treated the space charge as a small perturbation. This work then determined a space charge driven tune shift proportional to the current. Cousineau et al. studied the effects of envelope integer resonances numerically in~\cite{cou_lee_hol_dan_fed_03} in the context of the Proton Storage Ring, and showed that real lattices with piecewise continuous focussing elements exacerbate the issues presented by Sacherer.

To bypass these resonance based restrictions on beam current, Danilov and Nagaitsev proposed~\cite{nag_val_dan_10} and developed~\cite{dan_nag_10} the theory of nonlinear lattices as a method of storing bunches with large tune spreads. That these lattices are integrable means that particles at the design energy follow regular, nonchaotic trajectories in phase space, while off-momentum particles will undergo almost regular trajectories. Integrability insures that the particle trajectories are bounded in phase space, and the nonlinearity creates amplitude-dependent tunes in the beam, which prevents perturbations from resonantly driving single particles. Such tune spreads and the resulting nonlinear decoherence can enable intense particle beams to remain stable in the presence of undesired parametric resonances. It is also known~\cite{son_car_04} that for certain nonintegrable lattices, the dynamics will remain bounded and close to integrable, still generating strongly amplitude-dependent tune spreads. This method has already been shown to mitigate the formation of beam halo~\cite{son_car_05}.

In this paper, we demonstrate a number of examples of the effects of nonlinear decoherence. Because nonlinear decoherence is so important to the ideas here, we spend some time in Section II illustrating it with a model example. In Section III, we discuss the lattices under consideration and discuss the correct beam matching for them. We used the PyORBIT Python-wrapped version of the ORBIT tracking code~\cite{ost_hol_03, hol_dan_etc_02, shi_hol_dan_02, shi_hol_gor_09} to carry out these simulations, and give a brief overview in Section IV. The effects of nonlinear decoherence on the breathing modes generated by a beta mismatch in the beam is discussed in Section V, while in Section VI we demonstrate that the results in the previous sections lead to the prevention of halo formation by a mechanism discussed in~\cite{bruh_95}.

\section{Nonlinear Decoherence}

The key characteristic of the lattices proposed in~\cite{dan_nag_10, nag_val_dan_10, son_car_04} is the concept of \emph{nonlinear decoherence}, which arises from the frequency spread intrinsic to nonlinear systems. This is a phenomenon different from Landau damping~\cite{landau46}, a distinction which we will illustrate in this section. Specifically, Landau damping is an ensemble concept, whereas nonlinear decoherence is an intrinsically single-particle property.

Consider an example Hamiltonian which has a purely quartic potential:
\begin{equation}
H = \frac{p_x^2}{2} + \frac{1}{4} \lambda^4 x^4
\end{equation}
This Hamiltonian is completely integrable, and the Hamiltonian in terms of the action variable is
\begin{equation}
\overline{H} = \left ( \frac{ \lambda J_x}{2 \alpha} \right )^{4/3}
\end{equation}
where $\alpha  = \sqrt{\pi} \nicefrac{\Gamma(5/4)}{\Gamma(7/4)}$ is a constant. The tune for an oscillation with an action $J$ is therefore
\begin{equation}
\nu = \frac{1}{2 \pi} \frac{\partial \overline{H}}{\partial J} = \frac{1}{2 \pi} \frac{4}{3} \left ( \frac{\lambda}{2 \alpha} \right )^{4/3} J^{1/3}
\end{equation}

\begin{figure}
\includegraphics[scale=0.4]{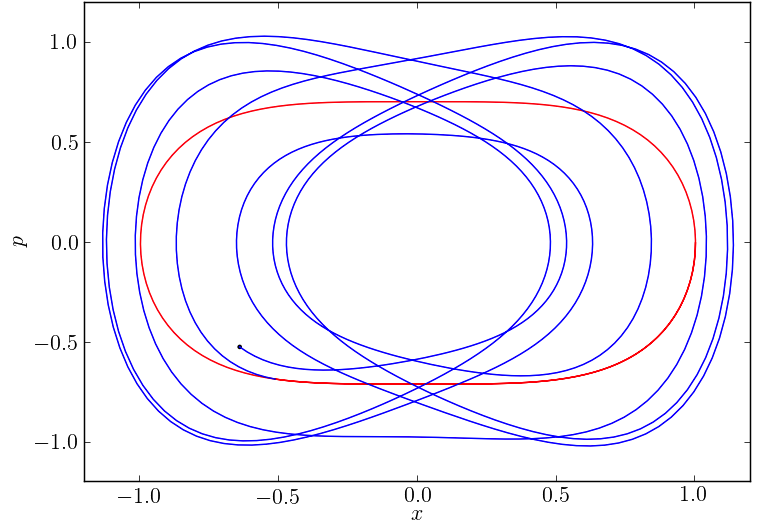}
\caption{The $p-x$ phase space trajectory of a quartic oscillator without resonant forcing (red) and with a forcing resonant with the initial frequency (blue).}
\label{quarticTrajectory}
\end{figure}

This Hamiltonian has an amplitude dependent frequency for closed trajectories in phase space, and so if any periodic forcing occurs at the initial frequency $\omega = \nicefrac{4}{3} \left ( \nicefrac{\lambda}{2 \alpha} \right )^{4/3} J^{1/3}$, after $J$ changes under the perturbation, it will no longer be resonant. This is illustrated in figure (\ref{quarticTrajectory}). The red trajectory represents the closed orbit before a $\sin (\omega_0 t)$ forcing begins. The blue trajectory is the resulting trajectory after the periodic perturbation begins. While the Hamiltonian is no longer integrable, the perturbation does not resonantly drive the particle to large amplitudes. 

For a harmonic oscillator Hamiltonian, which is the basis of linear accelerator lattices, the Hamiltonian in the action-angle variables is
\begin{equation}
\overline{H} = J \omega_0 
\end{equation}
for a frequency $\omega_0$, and hence the tune is independent of the amplitude. In this case, a periodic forcing would drive the particle to arbitrarily large amplitudes. This amplitude-dependent tune is the heart of nonlinear decoherence.

\begin{figure*}
\subfigure{\includegraphics[scale=0.4]{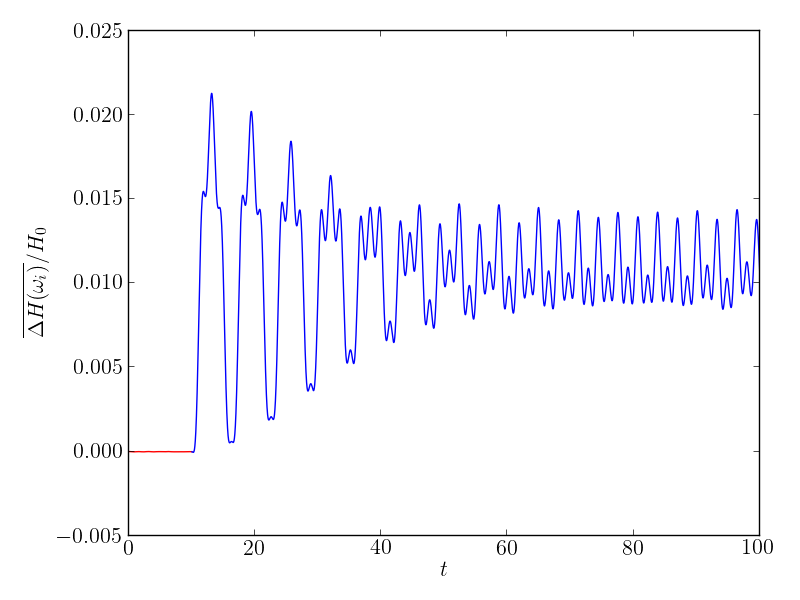}}
\subfigure{\includegraphics[scale=0.4]{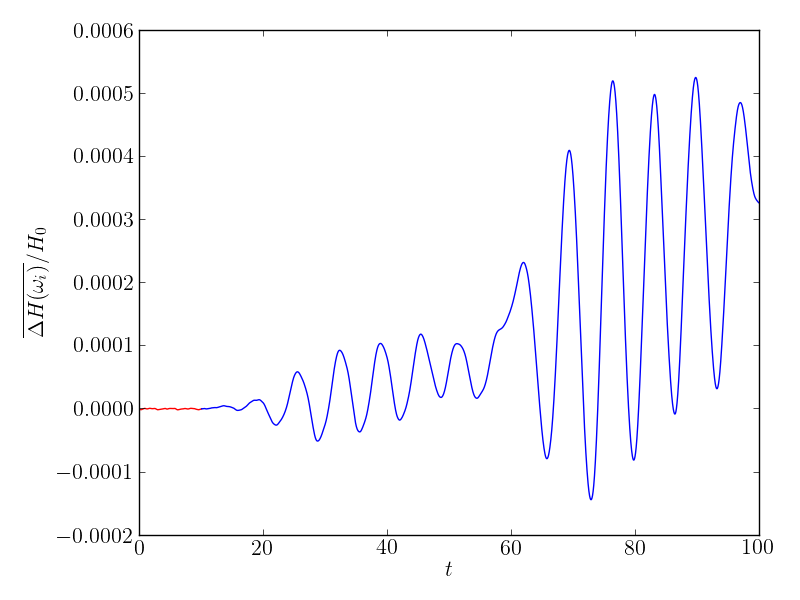}}
\caption{The average particle energy for the harmonic oscillator (left) and quartic oscillator (right) for many resonant forcing periods.}
\label{twoEnergies}
\end{figure*}

The effects of Landau damping and nonlinear decoherence are, in many practical ways, similar. To see this, consider the two plots in figure (\ref{twoEnergies}). We consider a 2D example, where each $H = H_x + H_y$, with anisotropic potential strengths. This allows us to create a matched ensemble of particles with an initial tune spread for the quartic oscillator.

To model Landau damping in the presence of a resonance, we integrated an ensemble of particles with a gaussian distribution of frequencies around some $\omega_0$ with a $5\%$ spread in both transverse directions. In this case, $\omega_x = 1.$ and $\omega_y = 2.$ and the forcing term drove the ensemble in each direction at the central frequency. The ensemble was matched to the Hamiltonian with a value $H_0 = 0.5$ in the way discussed later.

The quartic potential has $\lambda_x = 1$ and $\lambda_y = 2$, and is matched with $H_0 = 0.5$ as with the Landau damping example. The resonant forcing is set to the same strength as before, but with the forcing frequencies set to the average vertical and horizontal frequencies. In figure (\ref{twoEnergies}) we plot the average energy per particle added to the system by the resonant forcing.

For equal strengths in the perturbation, both prevent energy being added to the system on average. There is an important practical distinction between the Landau damping and nonlinear decoherence, though. Landau damping arises because an ensemble of particles has slightly different frequencies in a linear setting, and then after many oscillations resonant forcing stops adding energy to the ensemble as a whole (for a discussion of this, see e.g. Lee \cite{sylee} chap. 2, VIII or Chao and Zotter \cite{handbook} section 2.5.8). In terms of storage rings, Landau damping requires a spread in the betatron tunes, which in accelerator systems require some external dynamics. By contrast, nonlinear decoherence has no such requirement.

\begin{figure}
\includegraphics[scale=0.45]{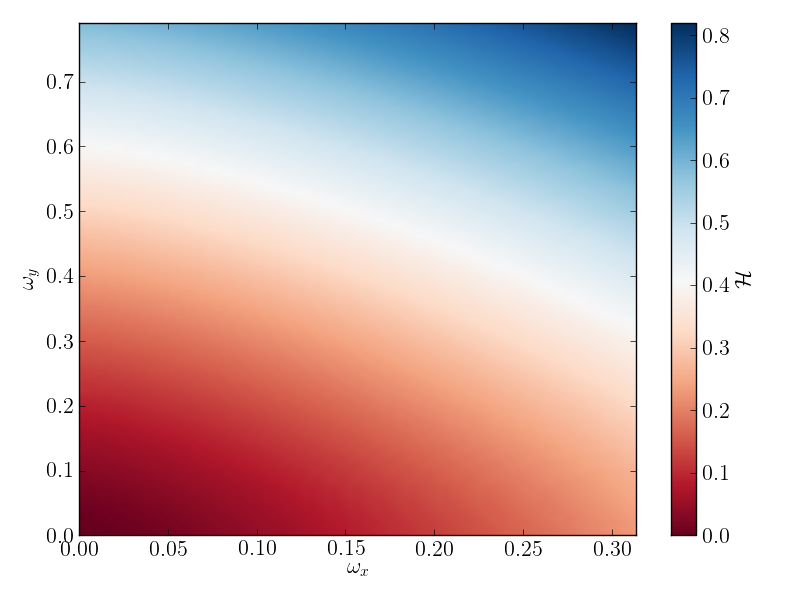}
\caption{Vertical and horizontal frequency as a function of the Hamiltonian for $\lambda_x = 1$ and $\lambda_y = 2$.}
\label{frequencySpread}
\end{figure}

An ensemble of particles with a fixed value of $H$ in the quartic potential will have a large intrinsic frequency spread, as illustrated in figure (\ref{frequencySpread}).
On this basis, a distribution of particles defined in a manner similar to the KV distribution, with
\begin{equation}
f(\vec{p}, \vec{q}) = \delta \left ( H(\vec{p}, \vec{q}) - \epsilon \right )
\end{equation}
will have a very large frequency spread. This distribution will be discussed further in section IV. For a harmonic oscillator Hamiltonian, this distribution will always have two frequencies, one for the vertical and the other for the horizontal, for all particles in the ensemble. In contrast, the quartic 2D oscillator has a broad frequency spread at fixed Hamiltonian.

In a storage ring setting, a large frequency spread at fixed value of the Hamiltonian that describes the single turn, single particle map, allows matched beams to exist with very large ranges in tune. This prevents coherent beam envelope oscillations as well as mitigating the effects of single particle resonances. 

Because the tune spread in linear lattices which arises from effects such as chromaticity may be limited by other dynamical considerations, it is clearly desirable to have nonlinear decoherence as a source of frequency spread. In existing linear lattice based machines, large energy spreads that cross resonances can lead to unstable operation, and sextupole corrections reduce the dynamic aperture. To keep the dynamic aperture larger, a more controlled method of producing predominately nonlinear lattices is required.

\section{Nonlinear Lattices}

In the previous section, we considered a Hamiltonian with a major drawback: it is not realizable in terms of magnetostatic fields. However, as Danilov and Nagaitsev showed in~\cite{dan_nag_10}, there do exist potentials which are realizable from beam optics, so long as the lattice is designed to have a section where the vertical and horizontal beta functions for the underlying linear lattice are equal.

For the purposes of these studies, we consider two of the lattices proposed in~\cite{dan_nag_10}: the integrable elliptic lattice and the chaotic bounded octupole lattice. We summarize these lattices here for clarity. To simplify notation, all coordinate discussions are in the normalized variables of the underlying linear lattice (the lattice for which the nonlinear potentials all have vanishing strength).

We consider first a lattice with integrable trajectories, taken from Sec. V.A of~\cite{dan_nag_10}, which we refer to as the integrable elliptic lattice (IEL). The IEL potential is derived from a simultaneous solution of Maxwell's equations with the Bertrand-Darboux equation~\cite{dar01, whit_37} for potentials which yield integrable motion in two dimensions with invariants quadratic in the canonical momentum. This lattice is related to linear strong focusing lattices, as it has completely integrable motion. A key distinction is that there is a strong action dependence in the frequency, which gives this lattice strong nonlinear decoherence.

\begin{figure}
\includegraphics[scale = 0.45]{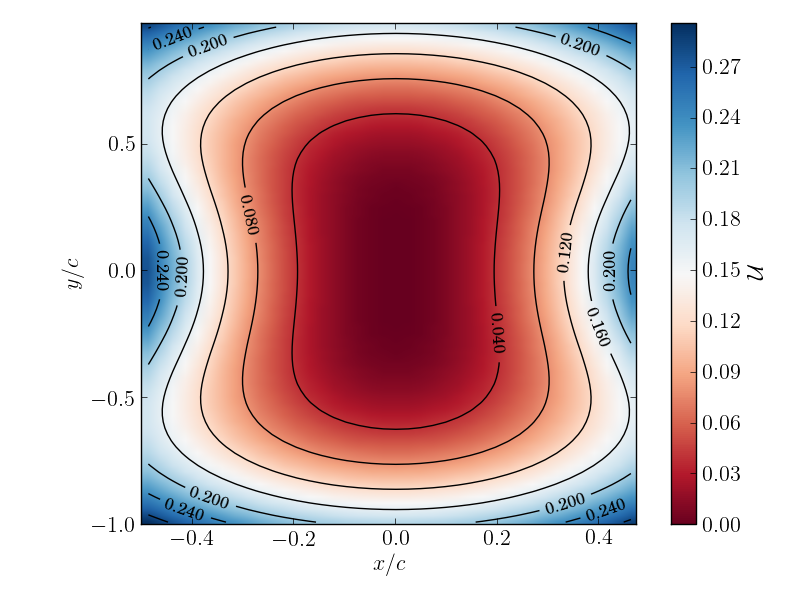}
\caption{Equipotential surfaces for the IEL potential.}
\label{hourglass}
\end{figure}

The resulting Hamiltonian, in normalized coordinates, is given by
\begin{equation}
H (p_x, p_y, x, y) = \frac{1}{2} \left ( p_x^2 + p_y^2 \right ) + \frac{1}{2} \left ( x^2 + y^2 \right ) + \frac{f_2(\xi) + g_2(\eta)}{\xi^2 - \eta^2}
\end{equation}
Here
\begin{equation}
\begin{split}
\xi = \frac{\sqrt{(x+c)^2 + y^2} + \sqrt{(x - c)^2 + y^2}}{2 c} \\
\eta = \frac{\sqrt{(x+c)^2 + y^2} - \sqrt{(x - c)^2 + y^2}}{2 c}
\end{split}
\end{equation}
are hyperbolic coordinates with foci at $x = \pm c$. The two potential functions are
\begin{equation}
\begin{split}
f_2(\xi) = \xi \sqrt{\xi^2 - 1} [ d + t \cosh^{-1}(\xi)]\\
g_2(\eta) = \eta \sqrt{1 - \eta^2} [ b + t \cos^{-1}(\eta)]
\end{split}
\end{equation}
where $d$, $b$, and $t$ are free parameters. For the purposes of this paper, we have taken $d = 0$, $b = \nicefrac{\pi}{2}$, and $t= -0.5$. By virtue of its time independence, the normalized Hamiltonian is an invariant of the motion. We show the equipotential surfaces for this Hamiltonian in figure (\ref{hourglass}). The importance of these equipotential surfaces will become clear when we discuss beam matching in the next section.

The other lattice considered here is taken from Sec. IV of~\cite{dan_nag_10}, which we refer to as the chaotic bounded octupole lattice (CBOL). The potential corresponds to a longitudinally varying octupole, with local strength that depends on the beta function of the underlying linear lattice. In normalized coordinates, the potential has the form:
\begin{equation}
U(x, y) = \frac{1}{2} \left ( x^2 + y^2 \right ) + \frac{\kappa}{4} \left ( x^4 + y^4 - 6 x^2 y^2 \right )
\end{equation}
This potential does not yield integrable motion because there is only one invariant of the motion (the Hamiltonian); however, it does have two important properties: (1) a large range of frequencies in the particle motion that depend upon the initial amplitude, and (2) trajectories which remain bounded, as long as the octupole strength is limited appropriately.

In linear lattices, beam matching is important to prevent rigid rotations in the normalized coordinate phase space, which manifests as periodic breathing modes in the transverse beam envelope. Because the oscillations are at twice the betatron frequency these can drive instabilities such as beam halo, as we will discuss later. These nonlinear lattices have more complicated dynamics, and thus the effects of beam mismatch are equally more complicated. The amplitude-dependent tune spread leads to filamentation for beams which are not properly matched, which can lead to an uncontrolled blow-up of the transverse beam size as a single-particle effect. To prevent this, we developed a generalization of conventional beam matching that reduces to the usual definition for linear lattices, and introduces a direct generalization of the concept of emittance to nonlinear lattices.

\section{Beam Matching}

\begin{figure*}
\subfigure[initial]{\includegraphics[scale=0.42]{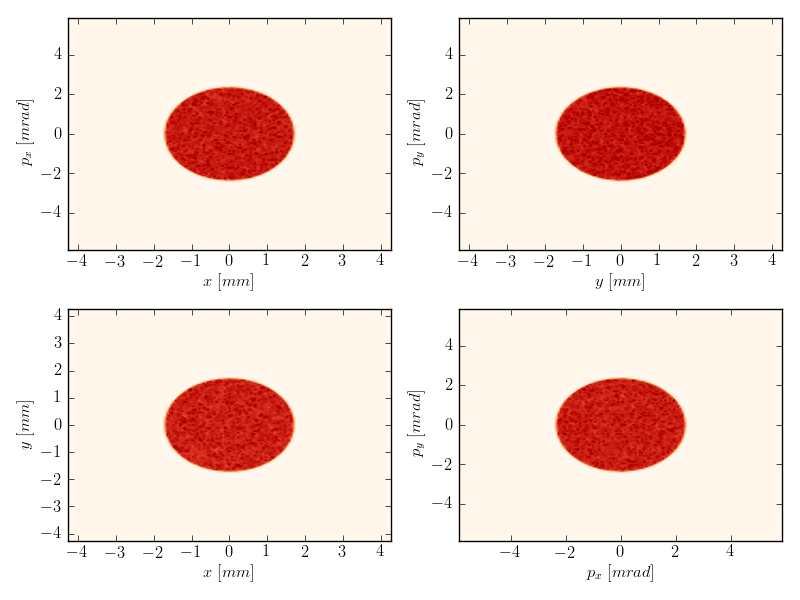}}
\subfigure[256 turns]{\includegraphics[scale=0.42]{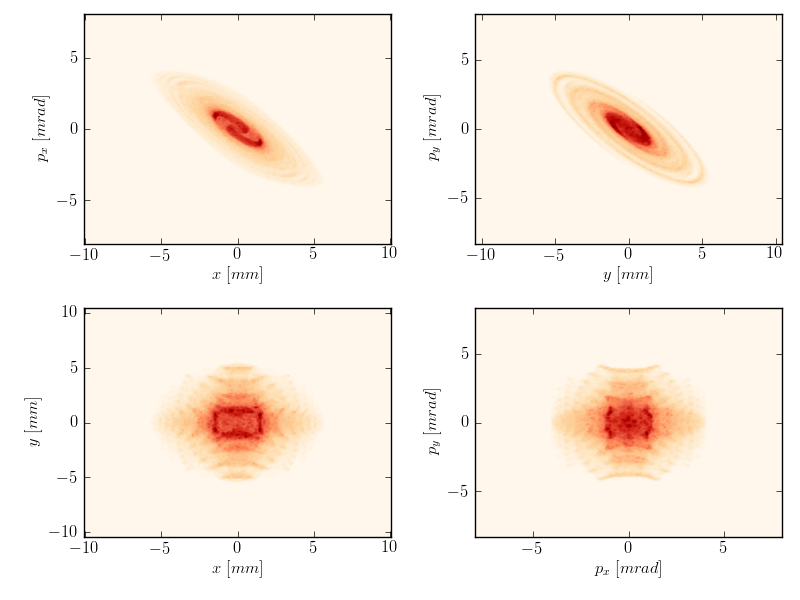}}
\subfigure[512 turns]{\includegraphics[scale=0.42]{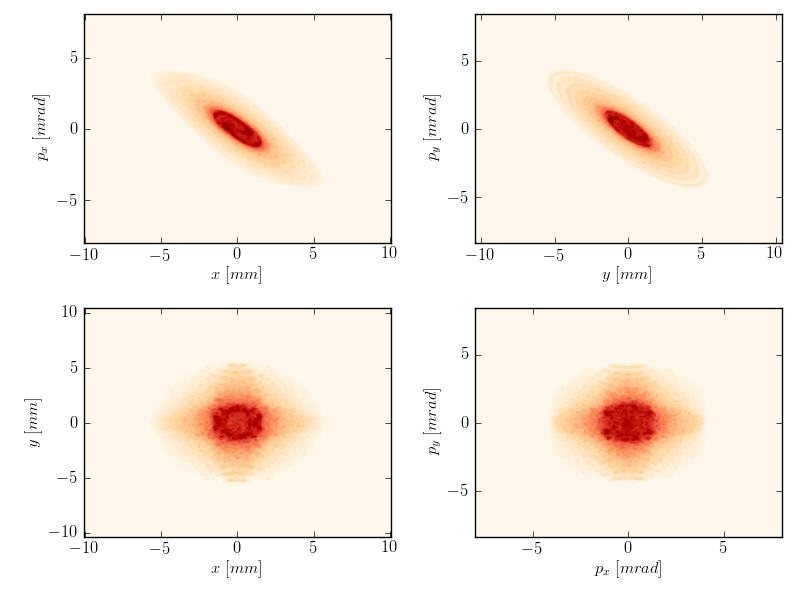}}
\subfigure[768 turns]{\includegraphics[scale=0.42]{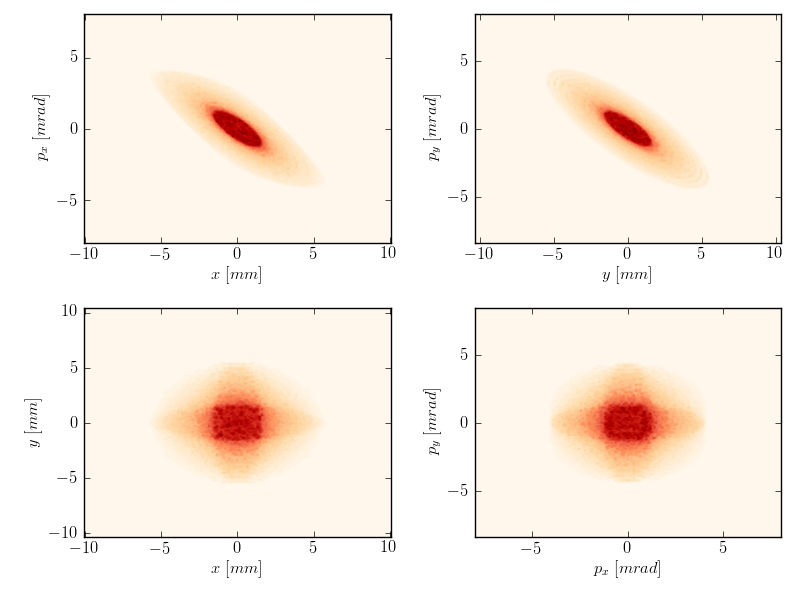}}
\caption{Evolution of a linearly matched beam in the integrable elliptic lattice. Note that the transverse size of the beam doubles in the $x$-$y$ plane.}
\label{ielMismatch}
\end{figure*}

Conceptually, single-particle beam matching is an effort to inject a beam with a phase space distribution that is stationary when measured at a fixed azimuthal point in a storage ring. Proper beam matching is extremely important for the practical implementation of the nonlinear lattices. Consider the plots in figure (\ref{ielMismatch}) as an example. From top left to bottom right are the initial beam and its evolution after the described number of passes through the IEL period. The initial beam is matched to the underlying linear lattice, when the elliptic potential strength is zero. There is, furthermore, no space charge or other collective forces.

Figure (\ref{ielMismatch}) shows that beam matching is important both theoretically and practically, as this mismatched IEL lattice sees the initial transverse beam size double after very few turns. To stably store a beam, as well as to interpret any results for studies in collective beam dynamics, we require a generalized concept of beam matching for these nonlinear lattices. Figure (\ref{ielMismatch}) also illustrates an important consequence of nonlinear decoherence -- an arbitrary initial distribution will filament in phase space until it reaches some static Vlasov equilibrium, in this case doing so after a few hundred iterations.

Any function of the invariants of motion form a stationary solution to the Vlasov equation \cite{nic_83, kt_86}. In 1958, Courant and Snyder derived their eponymous invariant \cite{cou_sny_58} for the piecewise linear magnetic fields that are used in every large accelerator today. The KV distribution is taken to be a delta function
\begin{equation}
f_{KV}(\vec{p}_N, \vec{q}_N) = \delta \left [ I(\vec{p}_N, \vec{q}_N) - \varepsilon_0 \right ]
\end{equation}
where $I$ is the Courant-Snyder invariant,
\begin{equation}
I = \gamma \vec{q}_N^2 + 2 \alpha \vec{q}_N \cdot \vec{p}_N + \beta \vec{p}_N^2
\end{equation}
which has the form of an harmonic oscillator Hamiltonian. Here $\alpha$, $\beta$, and $\gamma$ are the Twiss parameters, and $q_N$ and $p_N$ are the normalized transverse coordinate and momentum given by
\begin{equation}
\begin{array}{ccc}
\vec{q}_N = \vec{q}/\sqrt{\beta} & &
\vec{p}_N = \vec{p} \sqrt{\beta} - \beta' \vec{q}/(2 \sqrt{\beta})
\end{array}
\end{equation}
where here $q$ is either $x$ or $y$, and $p$ is either $p_x$ or $p_y$, respectively. The avenue to generalizing the above KV distribution would be to find another, similar function that reduces to the Courant-Snyder invariant for a purely linear lattice when the nonlinear potential strengths are zero.

Because the KV distribution is a Dirac delta function, and the invariants in a linear lattice are quadratic functions, it is straightforward to show that the $x-y$ phase space is a uniformly filled ellipse, which leads to linear space charge forces on all trajectories that remain in the distribution. As long as they are not too strong, these linear forces can be folded into the linear focusing forces from the external magnets and self-consistently included in the beam matching. Hence, the KV distribution with proper normalization of the Hamiltonian is a Vlasov equilibrium solution in the presence of finite beam current. As we explain below, our generalization of the KV distribution is strictly valid only in the zero current limit. Nevertheless, it shows a dramatic improvement on the problem illustrated in Figure (\ref{ielMismatch}).

By construction, the nonlinear lattices here have at least one invariant -- the Hamiltonian in the normalized coordinates. Thus, a distribution of the form
\begin{equation}
f\left ( \vec{p}, \vec{q} \right ) = \delta \left ( \mathcal{H} \left ( \vec{p}_N, \vec{q}_N \right ) - \epsilon_0 \right )
\end{equation}
is properly matched to this lattice. In this context, $\epsilon_0$ is the immediate generalization of emittance for the nonlinear lattices -- it parameterizes the volume of phase space occupied by the beam. It is interesting to note that because of the strong coupling between the transverse coordinates, it is no longer sensible to refer to a ``vertical" or ``horizontal" emittance for these lattices.

Like its linear counterpart, the generalized KV distribution uniformly fills its 2D projections -- relevantly for us in the $x-y$ projection it fills $U(x_N, y_N) = \epsilon_0$. However, because $U$ is not quadratic in the normalized coordinates, the curve is not an ellipse and the resulting space charge forces are not linear. For this reason, the generalized KV distribution described here is not an equilibrium for finite beam current. If one follows the example of the usual KV distribution and folds the linear space charge forces into the linear focusing of the external magnets, the resulting generalized KV distribution will be relatively close to a Vlasov equilibrium, where 'close' depends upon the strength of the space charge nonlinearities.

A linear lattice yields unconfined trajectories if the linear focusing forces are increased beyond linear stability, or if the linear space charge forces for a high-current KV beam become too large. For the two nonlinear lattices considered here, stable trajectories require additional restrictions on the beam size, depending on the strength of the nonlinear elements.

Specifically for the octupole potential, there are four hyperbolic fixed points at $(x_N, y_N) = (\pm \sqrt{\nicefrac{1}{2 \kappa}}, \pm \sqrt{\nicefrac{1}{2 \kappa}})$. The transverse beam size, as determined from $U(x_N, y_N) = \epsilon_*$, is bounded above by
\begin{equation}
r_* = \left ( \frac{1 + \sqrt{1 - 4 \kappa \epsilon_0}}{\kappa} \right )^{\nicefrac{1}{2}}
\end{equation}
where $r_*^2 = x_N^2 + y_n^2$ defines some characteristic radius for the beam.

If the initial beam overlaps these hyperbolic fixed points, some of the trajectories will be unbounded. This suggests a relationship between the generalized emittance and the normalized octupole strength to keep the entire beam confined
\begin{equation}
\kappa \leq \frac{1}{4 \epsilon_*}
\end{equation}
Since the level of nonlinear decoherence is related to the strength of $\kappa$, this puts a practical limitation on the beam emittance for any specified level of nonlinear behavior. The IEL has an analogous restriction -- the initial beam must not overlap the poles of the potential. These poles correspond to the magnetic pole faces of the elliptic elements, however, and therefore lie inside the magnetic material.

Given that the matching leads to stable, stationary single-particle distributions, we now have developed a beam which the sort of frequency spread illustrated in figure (\ref{frequencySpread}) that leads to nonlinear decoherence. This prevents the sort of rigid envelope oscillations that arise in linear lattice beam mismatch and creates a large tune spread that suppresses parametric resonances.

\section{Halo Suppression}

To illustrate the use of these nonlinear lattices, we now consider a canonical intensity-dependent source of beam loss: halo formation. Beam halo can arise from the resonant interaction of the coherent breathing modes of a mismatched beam (the core) in a linear lattice with any particles outside the phase space volume of the core. We refer to these particles as a pre-halo, sampling them from the phase space of a hypothetical matched distribution with the same emittance as the mismatched core. As the mismatched core's transverse beam envelope breathes at twice the betatron frequency, it drives the pre-halo particles resonantly in a linear lattice, pushing them to large amplitudes where the space charge force becomes nonlinear enough to no longer drive the particles on resonance. This effect requires two things: coherent beam envelope oscillations and a single frequency to characterize all the particle dynamics in a particular direction. The nonlinear lattices feature neither of these.

This particle-core model of beam halo formation was originally developed by O'Connell et al.~\cite{owmc_93} and Gluckstern~\cite{gluc_94}, and it has since been further analyzed and explored by many authors. It was shown by Bruhwiler~\cite{bruh_95}, for sufficiently strong mismatch, even for arbitrarily small space charge forces, that a pre-halo of particles nearby in phase space could be deterministically driven out into the halo. We present simulations of this parameter regime below.

For our simulations the underlying linear lattice, when the nonlinear strength is set to zero, has a betatron tune of $0.3$ in both transverse directions. The beam is a $100 A$, $1$ GeV cw proton beam with a generalized emittance of $\epsilon = 10^{-6} \mu m$. In every case, we matched the particle distribution to a beta function thirty per cent larger than the lattice beta function. The pre-halo constitutes $1 \%$ of the total beam current, and we consider halo particles as being any particle outside of two RMS beam radius. For the linear KV distribution, this is the sharp edge of the beam envelope, while for the IEL and CBOL lattices, this distinction is less clear. In all of the following plots, blue dots represent halo particles in this statistical definition, while the orange histogram represents the beam core.

\begin{figure*}
\subfigure[Blue dots indicate pre-halo particles outside of 2 RMS beam radius]{
\includegraphics[scale=0.4]{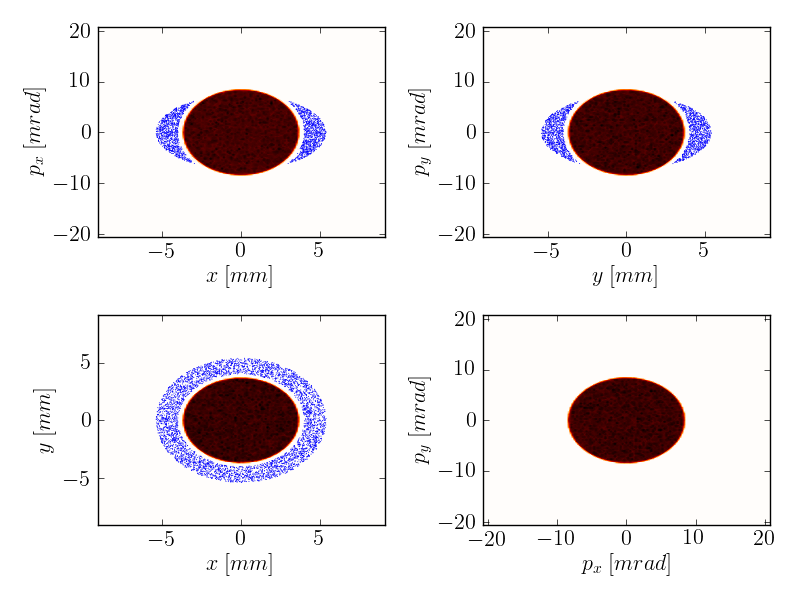}}
\subfigure[Particles in the pre-halo begin making large amplitude oscillations driven by the resonant space charge forcing of the beam core]{
\includegraphics[scale=0.4]{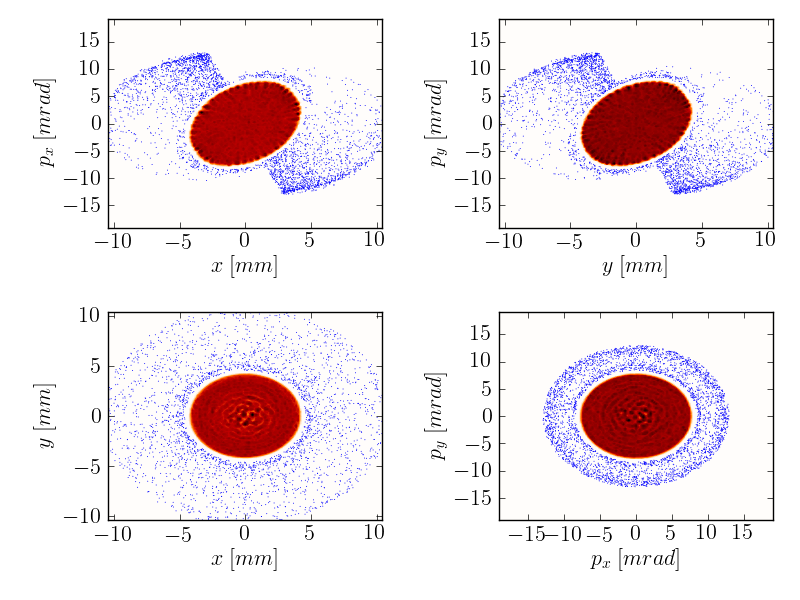}}
\caption{Coherent linear oscillations of the beam envelope drive halo formation.}
\label{lin_halo}
\end{figure*}

As described above, the rigid rotations of the beam core in phase space leads to the breathing modes in the transverse $x-y$ plane, which resonantly drives the pre-halo particles out into the halo. Very quickly, the particles get swept out to twice the beam radius, where the nonlinear space charge forces detune the forcing from resonance and the particles undergo stable orbits at large radii, as predicted. The lattice considered here is equivalent to the IEL and CBOL lattices with the extended nonlinear elements replaced by drifts.

This is the benchmark case -- a configuration rigged to generate beam halo rapidly, as illustrated in figure (\ref{lin_halo}). As has been pointed out before by Batygin~\cite{batygin}, much of this can be mitigated with intelligent beam matching to cancel the space charge effects. Here we examine the alternative route presented by nonlinear decoherence in the lattice proper.

\begin{figure*}
\subfigure[Blue dots indicate pre-halo particles outside of 2 RMS beam radius]{
\includegraphics[scale=0.4]{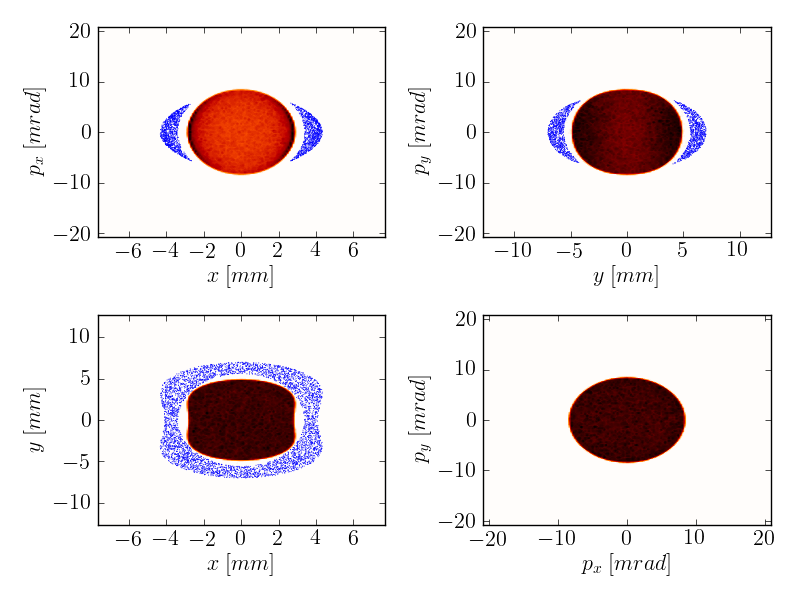}}
\subfigure[The IEL beam rapidly equilibrates to the properly matched physical dimensions with no halo formation]{
\includegraphics[scale=0.4]{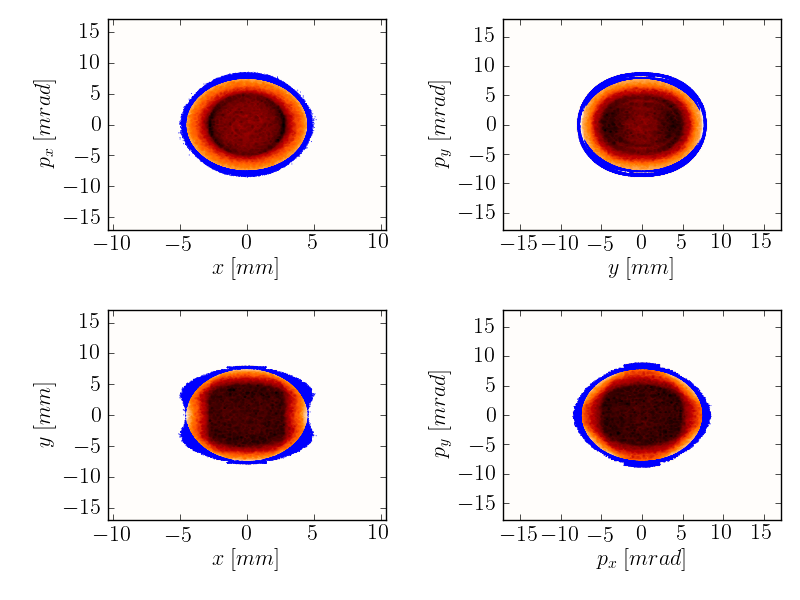}}
\caption{Nonlinear decoherence prevents the rapid formation of beam halo in the IEL.}
\label{iel_halo}
\end{figure*}

In both the IEL and CBOL cases, halo formation is completely mitigated. In the IEL, the system rapidly equilibrates to a configuration not dissimilar from the single-particle matched distribution outlined by the blue dots in figure (\ref{iel_halo}.a). The rigid core rotations in phase space present in the linear lattice are replaced by phase space filamentation for the IEL, which prevents any sort of parametric resonance from forming.

\begin{figure*}
\subfigure[Blue dots indicate pre-halo particles outside of 2 RMS beam radius]{
\includegraphics[scale=0.4]{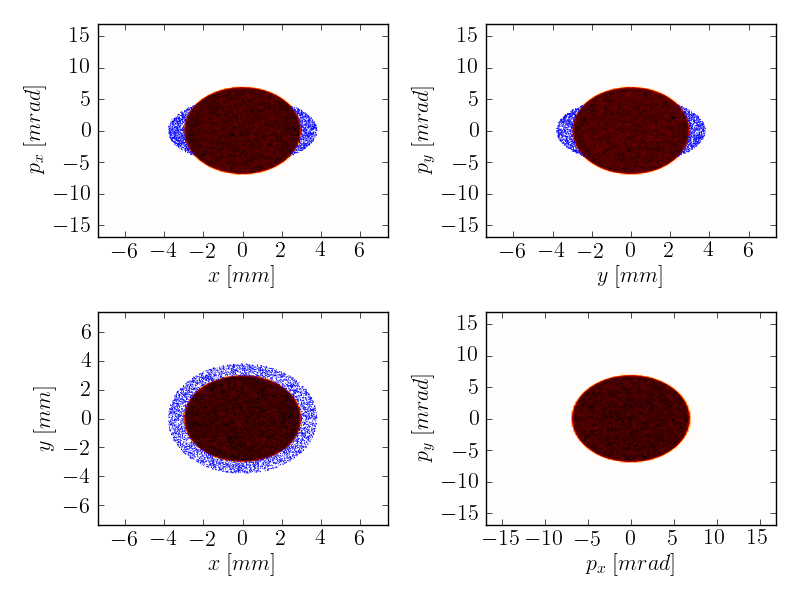}}
\subfigure[The CBOL beam rapidly equilibrates to the matched physical dimensions with no halo formation]{
\includegraphics[scale=0.4]{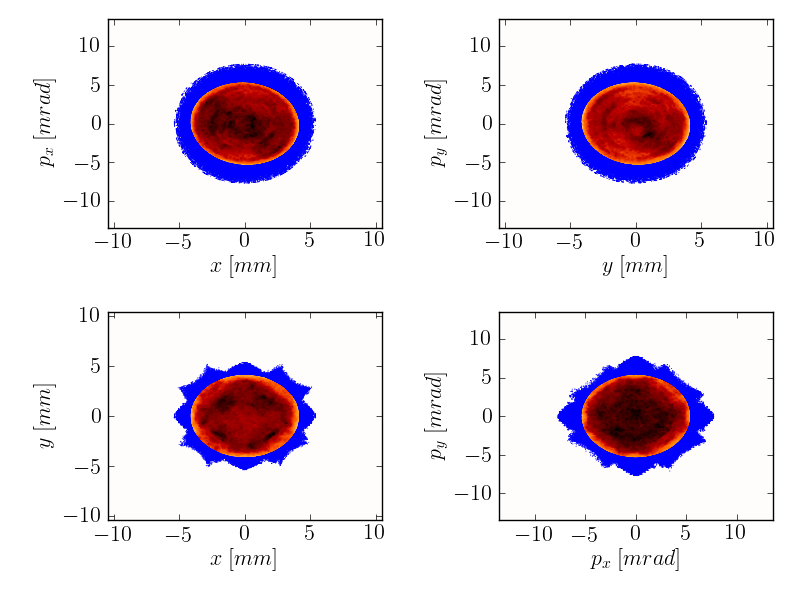}}
\caption{Nonlinear decoherence prevents the rapid formation of beam halo in the CBOL.}
\label{cbol_halo}
\end{figure*}

In the CBOL lattice, figure (\ref{cbol_halo}), the space charge forces along with the mismatch lead to more interesting shapes in the transverse $x$-$y$ plane due to the space charge forces, which contain all $e^{i 4 n \theta}$ harmonics, which appears in the structure of the outcroppings at the edge of the beam envelope. However, the nonlinear decoherence still suppresses halo formation.

The halo suppression seen in figures (\ref{iel_halo}) and (\ref{cbol_halo}) is an important development, and it shows the importance of studying this novel approach to beamline design for high intensity storage rings further. As noted in equation (10) above, there are an infinite number of Vlasov equilibria for for transverse particles distributions in a nonlinear lattice with at least one invariant of the motion (a 4D phase space with an invariant Hamiltonian, for example). It can be seen that an initial distribution which deviates significantly from any such Vlasov equilibrium (i.e. a beam core with 30\% mismatch, plus 1\% of the particles placed in a pre-halo) will evolve quickly towards a Vlasov equilibrium without pumping energy into halo formation. Simulations have shown this equilibrium to persist stably for a million iterations through the lattice. This property is extremely attractive, given that mismatch and deviation from design parameters are common for real accelerators and particle beams.

\section{Conclusion}

In both the integrable elliptic lattice and the chaotic bounded octupole lattice, the strong nonlinearities in the lattice play two roles. The primary role is to prevent rigid oscillations of the transverse beam envelope -- mismatch due to either injection errors or space charge causes phase space filamentation instead of rigid rotations. This prevents the parametric resonance behavior the envelope oscillations in the linear lattice drives. We speculate given these results that it would be possible to store a beam until the space charge forces became comparable to the focusing strengths, as the strong nonlinear decoherence inherent in either of these lattices will prevent any sort of resonant forcing to the entire ensemble of particles, even without longitudinal momentum spread and the resulting chromatic tune spread.

We have presented preliminary results for the efficacy of a novel lattice design which uses controlled nonlinearities to maintain the dynamic aperture while introducing nonlinear decoherence to the single-particle dynamics. It is this effect which efficiently prevents a variety of resonances -- here we have discussed the resonant interaction of beam mismatch oscillations and space charge to produce a beam halo. We have shown that these lattices are capable of completely preventing the formation of beam halo where a linear lattice would see the halo instability immediately. This is a promising initial result for future advances in the intensity frontier.

\begin{acknowledgments}
This work was funded by the US DOE Office of Science, Office of High Energy Physics under grant No. DE-SC0006247.
\end{acknowledgments}

\begin{appendix}
\section{PyORBIT}

The simulations were carried out using the PyORBIT tracking code maintained by Oak Ridge National Laboratory and freely available \footnote{The PyORBIT source code is maintained under the MIT license at Google Code \texttt{http://code.google.com/p/py-orbit/}} to the community.

The PyORBIT is a Particle-In-Cell accelerator simulation code, and it is a further development of the original ORBIT code~\cite{hcdh03}, which has been very useful for the SNS ring design and in simulations of collective effects for SNS and other projects around the world~\cite{scdg06}. The new code, like the original ORBIT, has a two-level structure. Time consuming calculations are performed on C++, and a high level simulation flow control is implemented in a scripting language. In PyORBIT the outdated and unsupported Super Code~\cite{han94} is replaced by Python, an interpreted, interactive, object-oriented, extensible programming language. We are not in the process of porting the ORBIT modules to PyORBIT.

At the present moment, the core of the PyORBIT code includes a Bunch class as a container for macro-particle coordinates and parameters, the TEAPOT-like~\cite{tal_sch_86} elements library and lattice classes to simulate rings and beam transport lines, a set of space charge modules, nonlinear lattice elements for integrable optics, collimation and foil injection models, the MPI library python shell, and a linear accelerator model. In PyORBIT the ring or transport line accelerator lattices can be constructed by using MAD 8 or SAD input files or directly in the python script. The linear accelerator structure is initialized from a PyORBIT specific XML file. The PyORBIT space charge modules have 2D, 2.5D (with possible perfect conducting wall boundary conditions), and 3D Poisson solvers based on the Fourier convolution theorem and discrete transformation (FFT) and the method of a image charge forces calculation suggested in~\cite{jon00}. The existing space charge modules have a low scalability for parallel calculations, and the development of new methods and relocation of the existing original ORBIT methods are underway.
\end{appendix}

\bibliography{hilat.bib}

\end{document}